\documentstyle[aps,prb,epsfig,multicol]{revtex}
\begin{document}
\title{The effect of compression on the global optimization of atomic clusters}
\author{Jonathan P.~K.~Doye}
\address{University Chemical Laboratory, Lensfield Road, Cambridge CB2 1EW, UK}
\date{\today}
\maketitle
\begin{abstract}
Recently, Locatelli and Schoen proposed a transformation of the potential energy that 
aids the global optimization of Lennard-Jones clusters with non-icosahedral global minima. 
These cases are particularly difficult to optimize because the potential energy surface 
has a double funnel topography with the global minimum at the bottom of the narrower funnel.
Here we analyse the effect of this type of transformation on the topography of the potential energy surface.
The transformation, which physically corresponds to a compression of the cluster, firstly
reduces the number of stationary points on the potential energy surface.
Secondly, we show that for a 38-atom cluster with a face-centred-cubic global minimum 
the transformation causes the potential energy surface to become increasingly dominated by the 
funnel associated with the global minimum. The transformation has been incorporated in the 
basin-hopping algorithm using a two-phase approach.
\end{abstract}
\begin{multicols}{2}

\section{Introduction}

One of the most important types of global optimization problem, and one which is 
particularly of interest to chemical physicists, is the determination of the 
lowest energy configuration of a molecular system, such as a protein, a crystal or 
a cluster.\cite{WalesS99} However, such a task can be very difficult because of the large number
of minima that a potential energy surface (PES) can have---it is generally expected that
the number of minima of a system will increase exponentially with size.\cite{Still99}
Therefore, if applications to large systems with realistic descriptions of the 
interatomic interactions are to be feasible, it is necessary that efficient global
optimization algorithms, which scale well with system size, are developed. 

A key part of this development is understanding when and why an algorithm 
is likely to succeed or fail, because, as well as providing useful information about
the limitations of an algorithm, this physical insight might be 
utilised in the design of better algorithms. 
This is the motivation behind the current paper. Here, we analyse the reasons
for the success of a recent algorithm when applied to the global optimization 
of Lennard-Jones (LJ) clusters for some particularly difficult sizes. 

The global optimization of LJ clusters has probably become the most common
benchmark for configurational optimization problems.\cite{WalesS99,Wille00} 
Putative global minima have been obtained for all sizes up to 309 
atoms,\cite{Wille87,Northby87,Coleman,Xue94,Gomez94,Pillardy,Doye95c,Doye95d,Deaven96,WalesD97,Barron97,Leary97,Romero99,Leary99}
and up-to-date databases of these structures are maintained on the web.\cite{Web,Barronweb} 
There are two types of difficulty for the LJ cluster problem. First, there is the general increase
in the number of minima with cluster size.\cite{HoareM76,Tsai93b}
Second, on top of this effect there are size-specific effects related 
to the topography of the PES.\cite{Doye99f}

For most of the clusters the topography of the PES aids global optimization. 
There is a funnel\cite{Leopold,Bryngelson95} from the high-energy 
liquid-like clusters to the low energy minima with structures based up on 
the Mackay\cite{Mackay} icosahedra. When there is a dominant low-energy icosahedral minimum at the bottom of the funnel,
such as when complete Mackay icosahedra can be formed, global optimization is particularly easy.

\begin{center}
\begin{figure}
\vglue-0.4cm
\epsfig{figure=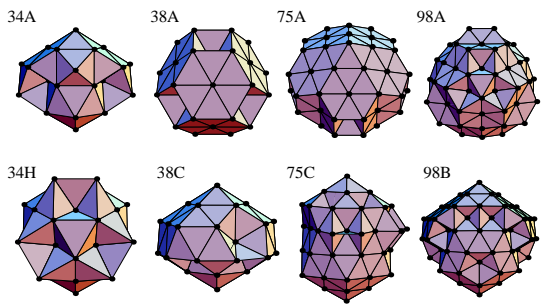,width=8.2cm}
\begin{minipage}{8.5cm}
\caption{\label{fig:structures}
The global minima and some low-lying minima of LJ$_{34}$, LJ$_{38}$, LJ$_{75}$ and LJ$_{98}$.
34A, 38C, 75C and 98B are based on Mackay icosahedra. 34H and 98A are Leary tetrahedra. 
38A is a face-centred-cubic truncated octahedron, and 75A is a Marks decahedron.
The letter gives the energetic rank of the minimum, i.e. global minima are labelled with an `A', etc.
}
\end{minipage}
\end{figure}
\end{center}

However, there are some sizes for which the global minimum is not icosahedral.
At $N$=38 the global minimum is a face-centred-cubic (fcc) truncated 
octahedron\cite{Gomez94,Pillardy,Doye95c} (38A in Figure \ref{fig:structures}),
at $N$=75--77 and 102--104 the global minima are based on Marks\cite{Marks84} decahedra\cite{Doye95c,Doye95d} 
(e.g.\ 75A in Figure \ref{fig:structures}), and at 
$N$=98 the global minimum is a Leary tetrahedron\cite{Leary99} (98A in Figure \ref{fig:structures}).
For these sizes the PES has a fundamentally different character. As well as the wide funnel
leading down to the low-energy icosahedral structures, there is a much narrower funnel which leads
down to the global minimum.\cite{Doye99f,Doye99c} 
Relaxation down the PES is much more likely to take the system into the wider funnel where it
is then trapped. The time scale for interfunnel equilibration is very slow\cite{Miller99b} because of the
large energy\cite{Doye99f} and free energy\cite{Doye99c} barriers between the two funnels.

As a result these eight clusters are hard to optimize, the larger examples being virtually impossible 
to optimize by traditional approaches, such as simulated annealing. 
However, these cases are solvable by a set of methods 
in which the `basin-hopping' transformation is applied to the PES.\cite{WalesD97} 
This transformation is used by the Monte Carlo minimization\cite{Li87a} or 
basin-hopping algorithm,\cite{WalesD97} and implicitly by all the most successful genetic 
algorithms.\cite{Deaven96,Deaven95,Gregurick,Niesse96a,Pullan97b,Wolf98,Barron99,Hartke99}
The transformation of the PES works by changing the thermodynamics of the 
clusters such that the system is now able to pass between the funnels more easily.\cite{Doye98a,Doye98e} 
However, the non-icosahedral global minima still take much longer to find than the
icosahedral global minima,\cite{WalesS99} and there is no way of knowing if one has waited long enough
to rule out the possibility of a non-icosahedral global minimum. This is illustrated by
the Leary tetrahedron at $N$=98. Despite the fact that powerful optimization
techniques had been applied to LJ$_{98}$,\cite{WalesD97,Barron99,Hartke99}
the global minimum was discovered only very recently.\cite{Leary99} 
Subsequently, it was confirmed that this minimum could be found by some of 
the previously applied methods.\cite{walespersonal,hartkepersonal} 

Given this background, it would be useful to develop techniques that are more efficient for 
these double-funnel examples. Two potential approaches have very recently been put forward. 
First, Hartke has achieved improvements in the genetic algorithm approach by forcing the 
system to maintain a diversity of structural types in the population, 
thus preventing the population becoming concentrated in
the icosahedral funnel.\cite{Hartke99}
Second, Schoen and Locatelli noted that the exceptions to the icosahedral structural motifs
are usually more spherical than the competing icosahedral structures. 
This is because the exceptions generally occur at sizes 
where both a particularly stable form for the alternative morphology is possible
and the icosahedral structures involve an incomplete overlayer.  
Therefore, Schoen and Locatelli added a term to the potential energy 
favouring compact clusters. 
Using this PES transformation, the non-icosahedral global minima at 
$N$=38, 98,102--104 were much more likely to be found by their multi-start 
minimization algorithm.\cite{Locatelli} 
An additional transformation had to be applied in order to find 
the global minima at $N$=75--77.

It is the reasons for the success of this second approach that we examine in this paper. 
In particular we show how Schoen and Locatelli's transformation affects the topography
of the PES. 
We also show how the transformation can be incorporated as an element of an 
existing algorithm, namely basin-hopping.

\section{Methods}

The atoms in the clusters interact via the Lennard-Jones potential:\cite{LJ}
\begin{equation}
E_{\rm LJ} = 4\epsilon \sum_{i<j}\left[ \left(\sigma\over r_{ij}\right)^{12} - \left
(\sigma\over r_{ij}\right)^{6}\right],
\end{equation}
where $\epsilon$ is the pair well depth and $2^{1/6}\sigma$ is the
equilibrium pair separation. To this, Schoen and Locatelli added a term 
proportional to $\sum_{i<j} r_{ij}$ which penalizes long pair distances.\cite{Locatelli}
Here, we use a slightly different form, which again acts to compress the cluster.
The energy for such a compressed Lennard-Jones (CLJ) cluster is given by 
\begin{equation}
E_{\rm CLJ}=E_{\rm LJ}+ \sum_i \mu_{\rm comp} {|{\bf r}_i-{\bf r}_{\rm c.o.m.}|^2\over \sigma^2},
\end{equation}
where $\mu_{\rm comp}$ is a parameter that determines the magnitude of the compression
acting on the cluster, and ${\bf r}_{\rm c.o.m}$ is the position of the centre of mass of the cluster.
We found the additional term to be approximately proportional to Schoen and Locatelli's expression,
and so the effect of the two transformations on the PES topography are virtually identical. 

To map the PES topography of these CLJ clusters we use the same methods as those
we have applied to LJ\cite{Doye99f,Doye99c} and Morse\cite{Miller99a} clusters 
to obtain large samples of connected minima and transition states that provide 
good representations of the low-energy regions of the PES. 
The approach involves repeated applications of eigenvector-following\cite{Cerjan} to 
find new transition states and the minima they connect.

In the basin-hopping algorithm,\cite{WalesD97,basinhopping} 
the transformed potential energy is given by 
\begin{equation}
 \tilde E({\bf x}) = {\rm min}\left\{ E({\bf x}) \right\},
\label{eq:BH}
\end{equation}
where ${\bf x}$ represents the vector of nuclear coordinates
and min signifies that an energy minimization is performed starting from ${\bf x}$.
Hence the energy at any point in configuration space is assigned to that of the local
minimum obtained by the minimization, and the transformed PES consists of
a set of plateaus or steps each corresponding to the basin of attraction surrounding a minimum on
the original PES.
This PES is then searched by constant temperature Monte Carlo.
Additionally, the algorithm has been found to be more efficient for clusters if the 
configuration is reset to that of the new local minimum at each accepted step.\cite{White98a}

There are two ways that one might incorporate a further PES transformation into this algorithm. 
One could use basin-hopping to first find the global minimum of the transformed PES, 
then reoptimize the $n_{\rm low}$ lowest energy minima under the original potential. 
However, if the global minimum of the original PES is not among 
the $n_{\rm low}$ lowest energy minima of the transformed PES this 
approach is bound to fail.

Alternatively, at each step one could first optimize a new configuration using 
the transformed potential, then reoptimize the resulting minimum using 
the original potential. By incorporating this second minimization the shortcomings
of the first approach are avoided. Furthermore, if the energy of this final minimum 
is used in the Metropolis acceptance criterion, 
the Boltzmann weight of each minimum is unchanged. However, the occupation 
probability of a particular minimum will be proportional the area of the basin 
of attraction of the minimum on the transformed rather than the original PES, i.e.\
\begin{equation}
\label{eq:2phase}
p_i\propto n_i \tilde A_i \exp(-\beta E_i), 
\end{equation}
where $n_i$ is the number of 
permutational isomers of $i$ and $\tilde A_i$ is the total area of the basins of attraction 
of the minima on $\tilde E$ which when reoptimized on $E$ lead to minimum $i$. 
Therefore, if the relative area of the global minimum
is larger on the transformed PES, optimization should be easier using this approach.
We refer to this version of the basin-hopping algorithm as 
two-phase basin hopping.
This variation is not much more computationally demanding than standard basin-hopping
because the starting point for the second minimization is likely to be close to
a minimum of the untransformed PES.

There is one further difference from previous implementations of the basin-hopping algorithm.
Previously, we had performed the minimization in Equation (\ref{eq:BH}) by 
conjugate gradient.\cite{Recipes} However, we have since found a limited memory
BFGS algorithm that is more efficient.\cite{Liu89}

\section{Results}

In global optimization the aim of transforming the potential energy surface is 
to make the global minimum easier to locate. Typically, one therefore wants the 
transformation to reduce the number of minima and the barriers between them. 
Furthermore, if the transformation is to change the relative energies of the minima, 
one wants the energetic bias towards the global minimum to increase. 

As the number of minima and transition states on the CLJ$_{13}$ PES 
is small enough that virtually all can be found, we can examine whether 
the compressive term has the first of the above effects by examining CLJ$_{13}$ 
as a function of $\mu_{\rm comp}$. 
The number of minima and transition states clearly decreases as 
$\mu_{\rm comp}$ increases (Table \ref{table:clj13}). 
It is interesting to note that minima with low symmetry preferentially disappear.
The PES transformation places the cluster in a harmonic potential about its 
centre of mass. This potential plays a role similar to a soft spherical box,
and so less compact minima disappear from the PES as $\mu_{\rm comp}$ increases. 
Similar results are found when periodic boundary conditions are applied---the 
number of minima is much less than for a LJ cluster of equivalent size 
and the number of minima decreases as the pressure in the cell is increased.\cite{Weber84,Heuer97}

It is also worth noting that the magnitude of the downhill barriers relative to the
energy difference between the minima decreases as $\mu_{\rm comp}$ increases (Table \ref{table:clj13}). 
In the terminology used by Berry and coworkers,\cite{Ball96} the profiles of the 
pathways to the global minimum become more staircase-like 
and less sawtooth-like with increasing $\mu_{\rm comp}$.
The combination of the changes to the number of stationary points 
and the barrier heights act to make relaxation to the icosahedral 
global minimum easier as the PES is further transformed.

Next, we examine the CLJ$_{38}$ cluster. For a cluster of this size it is not 
feasible to obtain a complete representation of the PES in terms of stationary points, 
so instead we obtain a good representation of the lower energy regions of the PES.
At each value of $\mu_{\rm comp}$ we obtained a sample of 6000 minima. The effect of 
$\mu_{\rm comp}$ on the number of stationary points, which we noted for CLJ$_{13}$, is 
again evident (Table \ref{table:clj38PES}). 
As $\mu_{\rm comp}$ increases, $n_{\rm search}$, the number of minima from which we have to 
perform transition state searches in order to generate the 6000 minima, 
increases and it becomes more likely that a new transition state does not connect 
to a new minimum, but rather to one already in our sample.

\begin{center}
\begin{figure}
\epsfig{figure=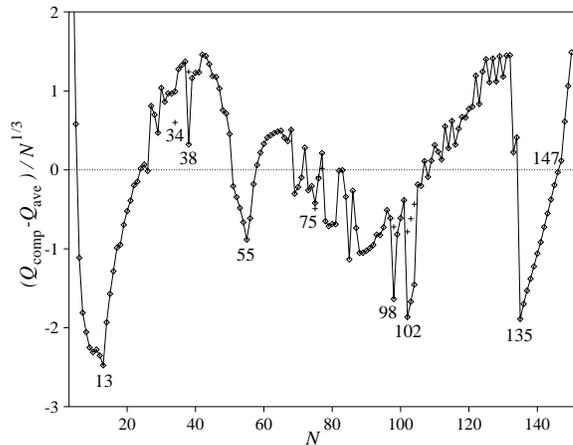,width=8.2cm}
\begin{minipage}{8.5cm}
\caption{\label{fig:qcomp_N}
$Q_{\rm comp}$ for the LJ$_N$ global minima.
To make the size-dependence more clear the zero is taken to be
the function, $Q_{\rm ave}$, a four paramater fit to the $Q_{\rm comp}$ values.
$Q_{\rm ave}=25.915 N -166.956 N^{2/3} +382.765 N^{1/3} - 293.972$.
Also included in the figure are isolated data points (crosses) corresponding to 
the non-global minima illustrated in Figure \ref{fig:structures} and the second lowest 
energy minima for $N$=76, 77 and 102--104.
}
\end{minipage}
\end{figure}
\end{center}

The second desired effect of a PES transformation is to 
change the energetics in a manner that makes the global minimum more favourable.
We can get a simple guide as to how the energies of the minima depend on $\mu_{\rm comp}$
if we assume there is no structural relaxation in response to changing $\mu_{\rm comp}$. Then 
$E_{\rm CLJ}=E_{\rm LJ}+ \mu_{\rm comp} Q_{\rm comp}$ 
where the order parameter, $Q_{\rm comp}=\sum_i |{\bf r}_i-{\bf r}_{\rm c.o.m.}|^2/\sigma^2$, is
evaluated at $\mu_{\rm comp}$=0.
From the values of $Q_{\rm comp}$ we can predict the changes in the 
relative energies of any two minima.

$Q_{\rm comp}$ is a measure of the compactness of the cluster, and from Figure \ref{fig:qcomp_N}
one can see how the compactness of the global minima depends on size. For the first two shells
the icosahedral global minima are most compact when complete Mackay icosahedra can be formed, 
e.g. $N$=13 and 55. 
However, for the third shell the most compact icosahedral structure is at $N$=135,
where twelve vertex atoms of the Mackay icosahedron are missing, rather than at $N$=147.

\begin{center}
\begin{figure}
\epsfig{figure=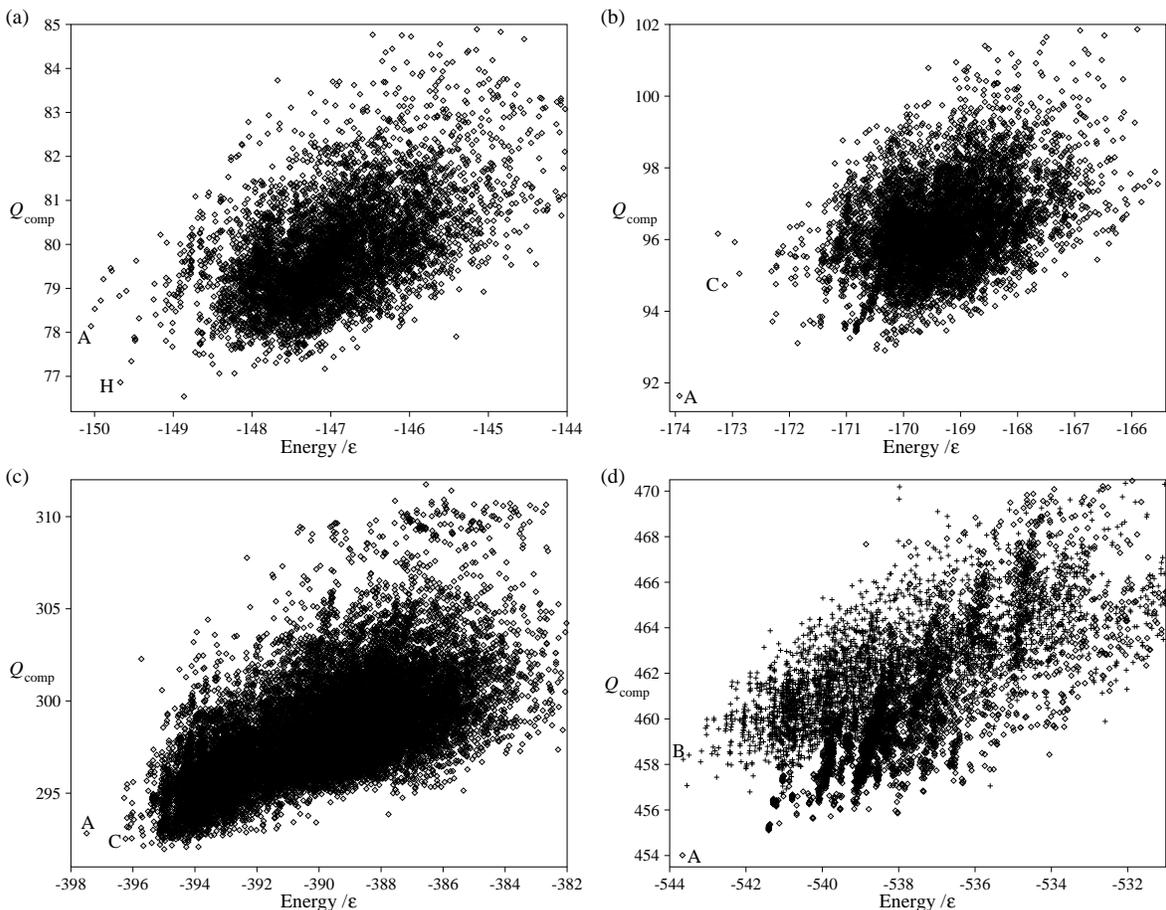,width=16cm}
\begin{minipage}{17.4cm}
\caption{\label{fig:qcomp}
Scatter plots of $Q_{\rm comp}$ against minimum energy for large samples of minima
for (a) LJ$_{34}$, (a) LJ$_{38}$, (a) LJ$_{75}$ and (d) LJ$_{98}$. 
The minima depicted in Figure \ref{fig:structures} are labelled by the letter
corresponding to their energetic rank.
In (d) there are two subsets: diamonds correspond to minima found when the search 
was started from the tetrahedral global minimum and crosses correspond to the set when started from 
the lowest energy icosahedral minimum.
There is no overlap between these two sets because no pathways connecting 
the two funnels were located.
The patterns of points for $Q_{\rm linear}=\sum_{i<j} r_{ij}/\sigma$ are
virtually identical to those of this figure, showing that the current transformation is 
effectively equivalent to Locatelli and Schoen's.
}
\end{minipage}
\end{figure}
\end{center}

\begin{center}
\begin{figure}
\epsfig{figure=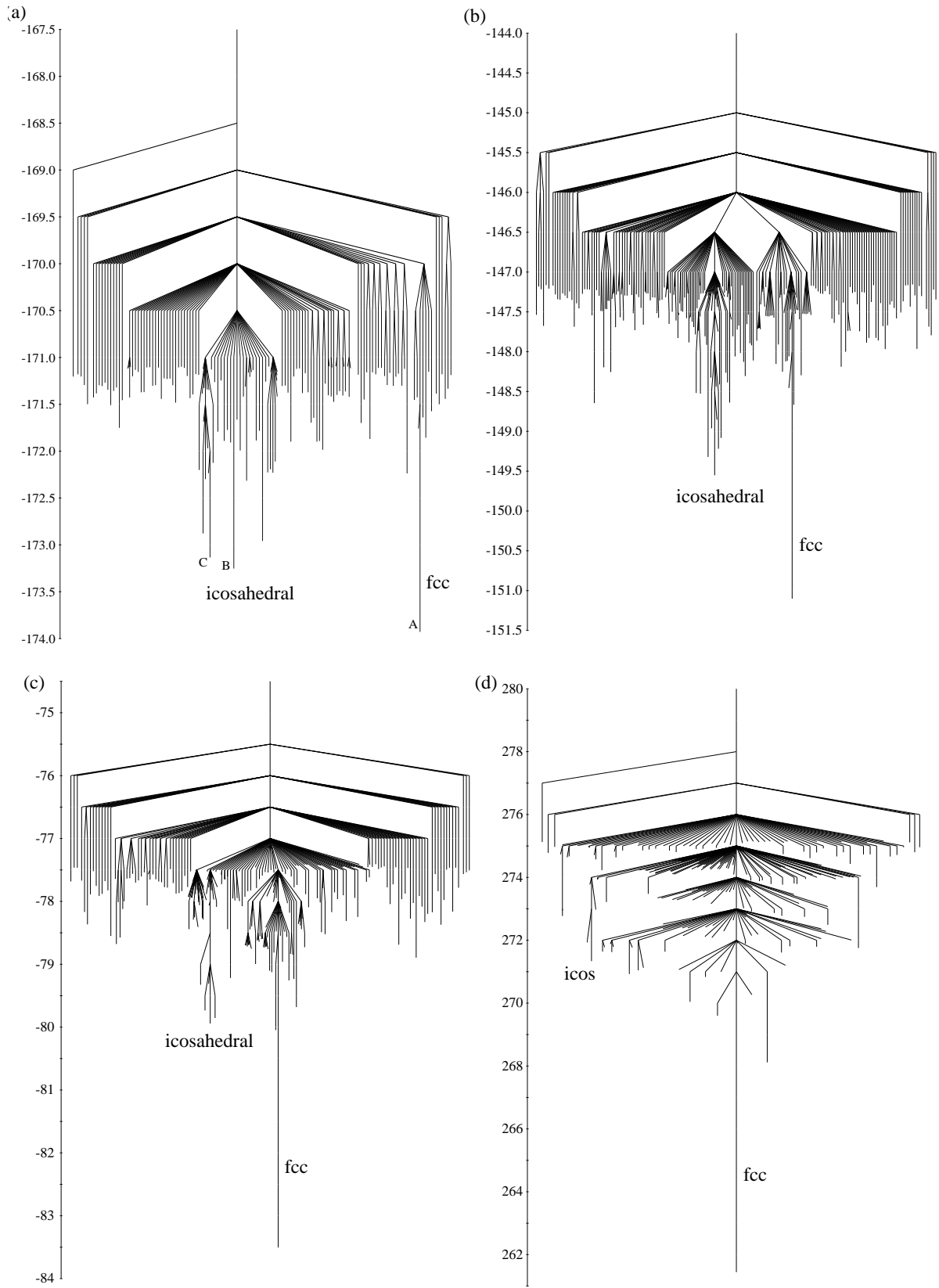,width=16cm}
\begin{minipage}{17.4cm}
\caption{\label{fig:trees}
Disconnectivity graphs of CLJ$_{38}$ for $\mu_{\rm comp}=$ 
(a) 0 (b) $0.25\epsilon$ (c) $1.0\epsilon$ and (d) $5\epsilon$. 
In (a) the 150 lowest-energy minima are represented in the graph, 
and in (b)--(d) the 250 lowest-energy minima are represented. 
The icosahedral and fcc funnels are labelled. 
The units of energy on the vertical axis are $\epsilon$.
}
\end{minipage}
\end{figure}
\end{center}

If we examine LJ$_{38}$ as an example of a cluster with a non-icosahedral global minimum, we 
see that this size corresponds to a pronounced minimum in Figure \ref{fig:qcomp_N}---the
truncated octahedron is particularly compact compared to the other global minima of similar size.
Furthermore, from Figure \ref{fig:qcomp}b we can see that the LJ$_{\rm 38}$ global minimum
has the lowest value of $Q_{\rm comp}$ of all the LJ$_{38}$ minima. 
Therefore, the energy gap between the global 
minimum and the lowest-energy icosahedral minimum increases with $\mu_{\rm comp}$ 
(Table \ref{table:clj38PES}). To visualize how this deepening of the fcc funnel
changes the PES topography we present disconnectivity graphs of CLJ$_{38}$ 
for a range of $\mu_{\rm comp}$ values in Figure \ref{fig:trees}.

Disconnectivity graphs provide a representation of the barriers between minima 
on a PES.\cite{Becker97,WalesMW98} 
In a disconnectivity graph, each line ends at the energy of a minimum. 
At a series of equally-spaced energy levels we compute which (sets of) minima are connected
by paths that never exceed that energy. 
We then join up the lines in the disconnectivity graph at the energy level where 
the corresponding (sets of) minima first become connected.
In a disconnectivity graph an ideal single-funnel PES would be represented by a 
single dominant stem associated with the global minimum to which the other minima 
directly join. For a multiple-funnel PES there would be a number of major stems which only join
at high energy. 

From the disconnectivity graph of LJ$_{38}$ one can deduce that the cluster has a 
double-funnel PES (Figure \ref{fig:trees}a). 
There is a narrow funnel associated with the global minimum, and a wider funnel associated
with the icosahedral minima. There are a number of low-energy minima at the bottom of the 
icosahedral funnel, which, although they have only small differences in the way the outer 
layer is arranged (e.g.\ the second lowest icosahedral minimum, 38C, is depicted in Figure 
\ref{fig:structures}), can be separated by moderate-sized barriers.
As a result there is a certain amount of fine structure at the bottom of the icosahedral funnel
with not all minima joined directly to the stem of the lowest-energy icosahedral minimum.
From the data in Table \ref{table:clj38PES} one can see that there are many more 
minima associated with the icosahedral funnel.

\end{multicols}

\begin{multicols}{2}

\begin{center}
\begin{figure}
\epsfig{figure=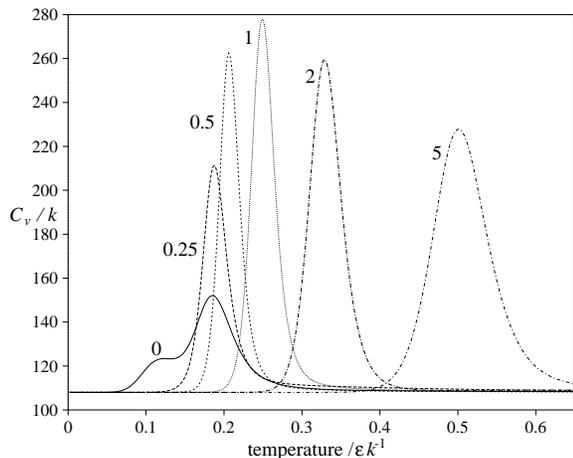,width=8.2cm}
\begin{minipage}{8.5cm}
\caption{\label{fig:cv} Heat capacity curves for CLJ$_{38}$ with the values of 
$\mu_{\rm comp}/\epsilon$, as labelled. The curves were calculated from our samples of 
6000 minima using the harmonic superposition method.\cite{Wales93a,Franke}
}
\end{minipage}
\end{figure}
\end{center}

As the fcc funnel becomes deeper with increasing $\mu_{\rm comp}$ it 
increases in size relative to the icosahedral funnel (Figure \ref{fig:trees}). 
By $\mu_{\rm comp}$=$5\epsilon$ the fcc funnel dominates the PES, 
and the disconnectivity graph has the form expected for an ideal single funnel 
with only a very small sub-funnel for the icosahedral minima. These
changes are also reflected in the number of minima associated with both funnels 
(Table \ref{table:clj38PES}).

These changes to the PES topography of course affect the thermodynamics. For LJ$_{38}$
there are two peaks in the heat capacity curve (Figure \ref{fig:cv}). 
The first is due to a transition from the fcc global minimum 
to the icosahedral minima, which is driven by the greater entropy of the latter. 
The second corresponds to melting. The first transition hinders 
global optimization because it is thermodynamically favourable for the cluster to enter
the icosahedral funnel on cooling from the molten state, where it can then be trapped.\cite{Doye98a,Doye98e}
However, as $\mu_{\rm comp}$ increases, the decreasing entropy of the icosahedral funnel can no longer
overcome the increasing energy difference between the global minimum and the icosahedral
funnel (Table \ref{table:clj38PES}) and so this first transition is suppressed. Consequently, the heat
capacity curves for the CLJ$_{38}$ clusters in Figure \ref{fig:cv} show only one peak, indicating
that the global minimum is most stable up to melting. 

Of course, the changes to the PES topography and thermodynamics mean that on relaxation down 
the PES the system is more likely to enter the fcc funnel as $\mu_{\rm comp}$ increases.
Furthermore, the energy barrier to escape from the icosahedral funnel relative to the energy
difference between the bottoms of the two funnels becomes smaller (Table \ref{table:clj38PES}), 
thus making escape from the icosahedral funnel easier.
To quantify these effects we performed annealing\cite{KirkSA} simulations for CLJ$_{38}$ 
at a number of values of $\mu_{\rm comp}$ (Table \ref{table:clj38anneal}).
For LJ$_{38}$ 80\% of the longer annealing runs ended at 
the bottom of the icosahedral funnel, and only 2\% at the global minimum.
However, by $\mu_{\rm comp}$=$5\epsilon$ 99.5\% of the long annealing runs reached the
global minimum.

\begin{center}
\begin{figure}
\epsfig{figure=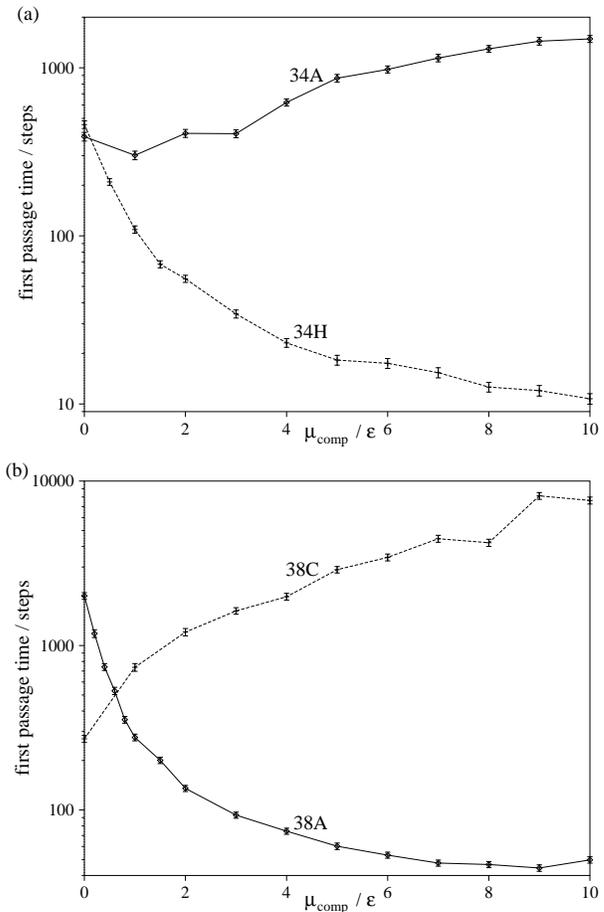,width=8.2cm}
\begin{minipage}{8.5cm}
\caption{\label{fig:fpt}
The $\mu_{\rm comp}$-dependence of the first-passage time (in MC steps) to find the specified minima of 
(a) LJ$_{34}$ and (a) LJ$_{38}$ from a random starting configuration in two-phase basin-hopping runs.
Each point represents an average over 400 runs. The temperature used is $1.0\epsilon k^{-1}$.
}
\end{minipage}
\end{figure}
\end{center}

Given the above, it is unsurprising that two-phase basin-hopping finds the global minimum
more rapidly as $\mu_{\rm comp}$ increases (Figure \ref{fig:fpt}b). At large $\mu_{\rm comp}$
the first-passage time is 40 times shorter than for LJ$_{38}$. Conversely, the first-passage time
to reach the icosahedral minimum 38C increases. These changes are driven by changes to 
$\tilde A_i$ in Equation (\ref{eq:2phase}). 
The basin of attraction of the global minimum increases in size relative to those of the icosahedral minima
as the PES is further transformed. 

Locatelli and Schoen's transformation works for LJ$_{38}$ because the global minimum is the most
compact spherical minimum. However, this does not necessarily have to be the case, even for 
those clusters with non-icosahedral global minima. 
From Figure \ref{fig:qcomp_N} one can see that the non-icosahedral global minima at $N$=98 and 102--104
have particularly low values of $Q_{\rm comp}$ and Figure \ref{fig:qcomp}d confirms that 
the Leary tetrahedron, 98A, has the lowest $Q_{\rm comp}$ value 
of all the LJ$_{98}$ minima. Therefore, Locatelli and Schoen were able to locate these global minima. 
However, for $N$=75-77 the values of $Q_{\rm comp}$ for the Marks decahedra
are not set apart from the nearby icosahedral global minima (Figure \ref{fig:qcomp_N})
and Figure \ref{fig:qcomp}c shows that there are a number of LJ$_{75}$ minima 
which have lower values of $Q_{\rm comp}$ than 75A. 
In particular, the icosahedral minimum 75C that is third lowest in energy has a lower $Q_{\rm comp}$, 
and the Marks decahedron is no longer the CLJ$_{75}$ global minimum beyond $\mu_{\rm comp}$=$3.1\epsilon$. 

The geometric root of this behaviour is that the Marks decahedra at $N$=75--77 are the least spherical
of the non-icosahedral global minima. The 75-atom Marks decahedron is somewhat oblate 
and some of the icosahedral minima with which 75A is competing are prolate by a similar degree, leading 
to comparable values of $Q_{\rm comp}$. Therefore, although the transformation may aid global 
optimization by reducing the number of minima and by increasing the energy of many minima relative
to the Marks decahedron, unlike for LJ$_{38}$ it does not remove the fundamental double-funnel character 
of the PES.  To locate the global minimum Locatelli and Schoen had to add an additional 
`diameter penalization' to the potential.\cite{Locatelli}

Locatelli and Schoen found that for many of the clusters their transformation 
did not aid global optimization. This was not unexpected, but simply reflects the 
fact that often the icosahedral global minima are not the most compact minima. 
We analyse one example. At $N$=34 it is possible to form a compact Leary tetrahedron (34H in 
Figure \ref{fig:structures}), which is the eighth lowest-energy LJ$_{34}$ minimum.
This structure has a significantly lower value of $Q_{\rm comp}$ than 
the global minimum (Figure \ref{fig:qcomp}a). 
As a result, the Leary tetrahedron becomes the CLJ$_{34}$ global minimum at 
$\mu_{\rm comp}$=$0.3\epsilon$. The results of two-phase basin-hopping runs 
are similar to those for LJ$_{38}$ in that as $\mu_{\rm comp}$ increases the compact 
non-icosahedral structure becomes significantly easier to 
locate and the low-energy icosahedral minima more difficult (Figure \ref{fig:fpt}). 
The difference, though, is that now this scenario is 
undesirable, because it is the global minimum that is becoming more difficult to reach. 

\section{Conclusions}

By analysing the effect of 
a compressive transformation on the PES topography
we have obtained insights into the reasons for its success in aiding the optimization of LJ clusters 
that have non-icosahedral global minima. Firstly, we have shown that the transformation reduces the number
of minima and transition states on the PES. Secondly, for examples where, as is often the case,
the non-icosahedral global minimum is the most compact structure, the transformation causes the funnel
of the global minimum to become increasingly dominant. For LJ$_{38}$ the PES has a double funnel, whilst at 
large $\mu_{\rm comp}$ the PES has an ideal single-funnel topography, enabling the system to relax easily 
down the PES to the global minima.
However, when as for LJ$_{75}$, the decahedral global minimum is only one of the more compact minima, 
the transformation is less beneficial for global optimization. 
By contrast, for sizes with icosahedral global minima the transition is often unhelpful, 
as we saw for LJ$_{34}$, because the global minimum is much less likely to be the most compact structure.
Therefore, the transformation needs to be used in combination with other methods. 
As the transformation is most likely to be successful for clusters where other methods fail 
it can act as a good complement to them.
For example, when the basin-hopping algorithm is applied, usually a series of runs are performed at each size. 
If one of the runs used the two-phase approach, this would increase the chance of success for 
those sizes where the PES had a multiple-funnel topography.

Other PES transformations could also be usefully employed alongside standard basin-hopping runs 
in this two-phase approach, if they are likely to aid global optimization for some sizes. 
For example, increasing the range of the potential is another transformation that
reduces the number of stationary points on the PES.\cite{Miller99a} 
Using the transformations alongside standard runs avoids one of the major difficulties 
associated with PES transformations. 
They are rarely universally effective, but rather there are likely to be some instances when they
destabilize the global minimum, thus making optimization more difficult. This is certainly the case 
when increasing the range of the potential, where the range-dependence 
of the most stable cluster structure is well-documented.\cite{Doye95c,Doye97d}

Although we have seen how a compressive transformation can be useful in aiding the global 
optimization of LJ clusters, an important question is how generally useful it will be. 
Although this question can only be definitively answered through applications to a 
variety of systems, one would expect it to be useful for metal and simple molecular clusters
that form compact structures, particularly those that favour 12-coordination. 
For these systems, as with LJ clusters, the strength of this approach would be locating those global minima 
that are not based on the dominant morphology, because the alternative morphologies are only 
likely to be most stable when they are compact and sherical.
It might also be useful in systems such as proteins where there are a large number of less compact 
unfolded configurations. However, it would not be useful for clusters of substances, 
such as water and silicon, which form open network structures where the liquid can be denser than
the solid.

\acknowledgements
J.P.K.D. is the Sir Alan Wilson Research Fellow at Emmanuel College, Cambridge.
The author is grateful to David Wales for supplying a modified version of the
basin-hopping code, and would also like to thank Marco Locatelli and Fabio Schoen 
for helpful discussions and for sharing results prior to publication.

\begin{center}
\begin{table}
\begin{minipage}{8.5cm}
\caption{\label{table:clj13}
The number of minima, $n_{\rm min}$, and transition states, $n_{\rm ts}$, for CLJ$_{13}$
as a function of $\mu_{\rm comp}$. 
For each minimum 30 transition state searches
were performed; these searches were parallel and antiparallel to the 
eigenvectors with the fifteen lowest eigenvalues.
$\overline{\Delta E}$, $\overline{b}_u$, $\overline{b}_d$ are the average energy difference, 
uphill barrier and downhill barrier, respectively, where the average 
is over all the non-degenerate rearrangement pathways. 
(Degenerate pathways connect different permutational isomers of the same minimum.)
$\overline{\Delta E}=\overline b_u -\overline b_d$.
}
\begin{tabular}{cccccccc}
 $\mu_{\rm comp}/\epsilon$ & 0  & 0.5 & 1   & 2.5 & 5   & 10  & 25  \\ 
\hline
$n_{\rm min}$ & 1467 & 769 &  470 &  169 &  75 &   33 &   10  \\   
$n_{\rm ts}$ & 12435 & 5820 & 3010 & 801 & 262 &  100 &   37 \\
$\overline{\Delta E}/\epsilon$ & 1.593 & 3.172 & 4.501 & 7.191 & 11.215 & 20.701 & 40.176 \\
$\overline{b}_{u}/\epsilon$   & 2.201 & 3.939 & 5.396 & 8.231 & 12.346 & 21.759 & 42.263 \\
$\overline{b}_{d}/\epsilon$ & 0.609 & 0.767 & 0.896 & 1.041 &  1.131 &  1.058 &  2.087 \\
$\overline{b}_{d}/\overline{\Delta E}$ & 0.382 & 0.242 & 0.199 & 0.145 & 0.101 & 0.051 & 0.052 \\
\end{tabular}
\end{minipage}
\end{table}
\end{center}

\begin{center}
\begin{table}
\begin{minipage}{8.5cm}
\caption{\label{table:clj38PES}
Properties of the CLJ$_{38}$ PES 
for a sample of 6000 connected minima as a function of $\mu_{\rm comp}$. 
$n_{\rm ts}$ is the number of transition states connecting these minima. 
$\Delta E$ is the energy difference between the global minimum and the lowest energy icosahedral minimum
and $b_{\rm fcc}$ ($b_{\rm icos}$) is the energy barrier that has to be overcome to 
escape from the fcc (icosahedral) funnel and enter the icosahedral (fcc) funnel.
Of course, $\Delta E=b_{\rm fcc}-b_{\rm icos}$.
For the $n_{\rm search}$ lowest-energy minima 20 transition state searches
were performed; these searches were parallel and antiparallel to the 
eigenvectors with the ten lowest eigenvalues.
$n_{\rm fcc}$ and $n_{\rm icos}$ are the numbers of minima in the fcc and icosahedral funnels
at the energy at which the two funnels become connected. 
}
\begin{tabular}{ccccccc}
$\mu_{\rm comp}/\epsilon$ & 0 & 0.25 & 0.5 & 1 & 2.5 & 5 \\
\hline
$n_{\rm ts}$   & 8633 & 9111 & 9911 & 11656 & 17137 & 23270 \\
$n_{\rm search}$ & 1271 & 1277 & 1491 & 1924 & 3107 & 4253 \\
$\Delta E$ & 0.676  & 1.550 & 2.274 & 3.564 & 6.120 &  9.893 \\
$b_{\rm fcc}/\epsilon$ & 4.219 & 4.795 & 5.256 & 6.143 & 8.892 & 12.659 \\
$b_{\rm icos}/\epsilon$ & 3.543 & 3.245 & 2.981 & 2.580 & 2.772 &  2.766 \\
$b_{\rm icos}/\Delta E$ & 9.893 & 2.094 & 1.311 & 0.724 & 0.453 & 0.280 \\
$n_{\rm fcc}$   & 92 & 113 & 73 & 106 & 104 & 86 \\
$n_{\rm icos}$  & 912 & 439 & 194 & 27 & 5 & 6 \\ 
$n_{\rm fcc}/n_{\rm icos}$ & 0.11 & 0.26 & 0.38 & 3.93 & 20.8 & 14.33 \\
\end{tabular}
\end{minipage}
\end{table}
\end{center}

\begin{center}
\begin{table}
\begin{minipage}{8.5cm}
\caption{\label{table:clj38anneal}
Results of annealing simulations for CLJ$_{38}$ as a function of $\mu_{\rm comp}$.
$f_{O_h}(n_{\rm cycles})$ is the fraction of the annealing runs that terminated at the global
minimum, and $f_{\rm icos}$ is the fraction of runs that ended in the lowest five icosahedral minimum.
Each annealing run involves a linear decrease in the temperature 
from the liquid to 0K in $n_{\rm cycles}$ Monte Carlo cycles. The results are averages over 200 annealing runs.
}
\begin{tabular}{ccccccc}
$\mu_{\rm comp}/\epsilon$ & 0 & 0.25 & 0.5 & 1 & 2.5 & 5 \\
\hline
$f_{O_h}(10^6)$ & 0\% &  2.5\% & 7.5\% & 19.5\% & 67\% & 79.5\% \\
$f_{O_h}(10^7)$ & 2\% & 14\% & 31\% & 66.5\% & 97\% & 99.5\% \\
$f_{\rm icos}(10^6)$ & 37\% & 29.5\% & 12.5\% & 7.5\% & 1\% & 0\% \\
$f_{\rm icos}(10^7)$ & 80\% & 56.5\% & 38\% & 6.5\% & 0\% & 0\% \\
\end{tabular}
\end{minipage}
\end{table}
\end{center}

\end{multicols}
\end{document}